\documentclass[twocolumn,english,showpacs,amsmath,amssymb,groupedaddress,superscriptaddress,aps,prl,floats]{revtex4}
\usepackage[]{fontenc}
\usepackage[latin1]{inputenc}
\usepackage{subfigure}
\usepackage{amsmath}
\usepackage{color}
\usepackage{graphicx}
\usepackage{amssymb}

\makeatletter
\usepackage{epsfig}

\usepackage{dcolumn}

\usepackage{bm}

\usepackage{amsfonts}

\usepackage{graphicx}%

\newcommand{\beq}{\begin{eqnarray}}

\newcommand{\eeq}{\end{eqnarray}}

\usepackage{babel}
\makeatother
\begin{document}

\title{Incommmensurability and unconventional superconductor to insulator
transition in the Hubbard model with bond-charge interaction}

\author{A. A.~Aligia}
\affiliation{Comisi\'on Nacional de Energ\'ia At\'omica, Centro At\'omico Bariloche and
Instituto Balseiro, 8400 S.C. de Bariloche, Argentina}

\author{A. Anfossi}
\affiliation{Dipartimento di Fisica del Politecnico and CNISM, corso Duca degli Abruzzi
24, I-10129, Torino, Italy}
\affiliation{BIFI, Universidad de Zaragoza, Corona de Arag\'on 42, 5009 Zaragoza, Spain}

\author{L. Arrachea}
\affiliation{Departamento de F\'{\i}sica de la Materia Condensada, Universidad
de Zaragoza,  5009 Zaragoza}
\affiliation{BIFI, Universidad de Zaragoza, Corona de Arag\'on 42, 5009 Zaragoza, Spain}

\author{C. Degli Esposti Boschi}
\affiliation{Unit\`{a} CNISM and Dipartimento di Fisica dell'Universit\`{a} di Bologna,
viale Berti-Pichat 6/2, I-40127, Bologna, Italy}

\author{A. O.\ Dobry}
\affiliation{Instituto de F\'{\i}sica Rosario, CONICET-UNR, Bv. 27 de
Febrero 210 bis, 2000 Rosario, Argentina.}

\author{C.\ Gazza}
\affiliation{Instituto de F\'{\i}sica Rosario, CONICET-UNR, Bv. 27 de
Febrero 210 bis, 2000 Rosario, Argentina.}

\author{A. Montorsi}
\affiliation{Dipartimento di Fisica del Politecnico and CNISM, corso Duca degli Abruzzi
24, I-10129, Torino, Italy}

\author{F. Ortolani}
\affiliation{Dipartimento di Fisica dell'Universit\`{a} di Bologna and INFN, viale
Berti-Pichat 6/2, I-40127, Bologna, Italy}

\author{M. E.\ Torio}
\affiliation{Instituto de F\'{\i}sica Rosario, CONICET-UNR, Bv. 27 de
Febrero 210 bis, 2000 Rosario, Argentina.}

\pacs{71.10.Fd,71.10.Hf,71.10.Pm,71.30.+h}
\date{\today}

\begin{abstract}
We determine the quantum phase diagram of the one-dimensional Hubbard
model with bond-charge interaction $X$ in addition to the usual Coulomb
repulsion $U>0$ at half-filling. For large enough $X<t$
the model shows three phases. For large $U$ the system is in
the spin-density wave phase as in the usual Hubbard
model. As $U$ decreases, there is first a spin transition to a spontaneously
dimerized bond-ordered wave phase and then a charge transition
to a novel phase in which the dominant correlations at large distances
correspond to an incommensurate singlet superconductor.
\end{abstract}

\maketitle

The Hubbard model has been originally proposed to describe the effect
of the Coulomb interaction in transition metals, which usually contain
localized orbitals. Other real compounds containing more extended
orbitals cannot in general be properly described by this simple Hamiltonian.
Well-known examples are several quasi-one-dimensional (1D) materials that
have been recently investigated \cite{1dmat}, which exhibit a variety
of phases that cannot be explained with the usual Hubbard model. Additional
interactions should be included. A natural interaction that arises
in systems with extended orbitals is the bond-charge interaction $X$
\cite{cam}. In fact, it is natural to assume that the charge
in the bond affects screening and the effective potential acting on
valence electrons, and therefore the extension of the Wannier orbitals
and the hopping between them should vary with the charge.

This leads to the $U-X$ Hamiltonian:
\begin{eqnarray}
H & = & -t\sum_{\sigma=\uparrow,\downarrow,\langle ij\rangle}(c_{i\sigma}^{\dagger}c_{j\sigma}+{\text H.c.})+U\sum_{i}n_{i\uparrow}n_{i\downarrow}\nonumber \\
 &  & -X\sum_{\sigma,\langle ij\rangle}(c_{i\sigma}^{\dagger}c_{j\sigma}+{\text H.c.})(n_{i-\sigma}
 +n_{j-\sigma}).\label{hamil}
\end{eqnarray}
This model has been studied in two dimensions, motivated by a theory
of hole superconductivity \cite{hirsch}. A modified version of it
has been derived as an effective model for the cuprates and shows
enhanced $d$-wave superconducting correlations \cite{lili}. Recently,
this model has been paramount to broader audiences, and its relevance
has been discussed in the context of mesoscopic transport \cite{impu}
and quantum information \cite{entro,agm}.

In 1D, there are bosonization \cite{japa,tll} and numerical \cite{tll}
results available. However, at half-filling, the effect of $X$ disappears
in the standard bosonization treatment and a behavior different from
the usual Hubbard model was not expected in these studies. For
$X=t$, an exact solution is available \cite{exac}.
In this case the ground state is highly degenerate: the transition to a metallic state takes place at $U_c= 4 t>0$, but the response of the
system to an applied magnetic flux indicates that it is not superconducting
\cite{flux}. In view of the previous studies, the recent evidence
of an insulator-metal transition driven by $X<t$ at finite $U_c>0$ at
half-filling comes as a surprise \cite{sup-ins}. The nature of the
metallic phase and the character of the transition have not been
fully elucidated, though the possibility of superconductivity has been suggested.

In this Letter we employ several analytical and numerical techniques
to calculate accurately the phase diagram of the model at half-filling
in 1D and to determine the nature of each phase. We establish that
the insulator-metal transition is of commensurate-incommensurate
(CIC) type to a phase with dominating singlet superconducting (SS)
correlations.
Remarkably, unlike other CIC transitions \cite{ttpu,igor}, it
is not driven by one-body effects like chemical potential or the emergence of more than two Fermi points in the noninteracting dispersion relation,
but by strong correlations induced by large enough $X$.
In addition, we unveil that inside
the insulating phase there is a spin transition separating the
expected spin-density wave (SDW) for $U>U_{s}$ from a spontaneously
dimerized bond-ordered wave (BOW) phase
for $U_c<U<U_{s}$. This transition is of Kosterlitz-Thouless (KT) type
and a spin gap opens in the BOW phase.

The nature of each phase and the qualitative aspects of the phase
diagram can be understood by a weak coupling bosonization analysis
\emph{provided} it includes vertex
corrections of second order in $X$ to the coupling constants
and one term of order $a^{2}$ in the bosonization of the bond-charge interaction
as described below, where $a$ is the lattice constant.
A bosonized version of (\ref{hamil})
is given by the following Hamiltonian density:
\begin{eqnarray}
{\mathcal{H}} & = & {\mathcal{H}}_{\sigma}^{0}+{\mathcal{H}}_{\rho}^{0}
 +  \frac{2g_{1\perp}}{(2\pi\alpha)^{2}}\cos(\sqrt{8}\phi_{\sigma})-
\frac{2g_{3\perp}}{(2\pi\alpha)^{2}}\cos(\sqrt{8}\phi_{\rho})\nonumber\\&+&
\frac{2 g_{\sigma\rho}}{(2\pi \alpha)^2}
\cos(\sqrt{8}\phi_{\sigma})\partial_{x}\phi_{\rho}
\label{hboso},
\end{eqnarray}
where ${\mathcal{H}}_{\sigma}^{0}$ and ${\mathcal{H}}_{\rho}^{0}$
are the usual known quadratic forms and $\alpha$ is a short range cutoff in
the bosonization procedure.
The first line of (\ref{hboso}) has the structure of the previously studied
bosonized theory \cite{japa}, which corresponds to two
decoupled sine-Gordon field theories, one for the spin ($\phi_{\sigma}$)
and the other for the charge ($\phi_{\rho}$).
In order to take into account the effect of the bond-charge interaction on the phase diagram of the system, we included vertex corrections of second
order in $X$ in the definition of the the coupling
constants $g_{i}$, due to virtual processes involving states far from
the Fermi energy \cite{japan}. In addition, we took into account the
usually neglected  $g_{\sigma\rho}$ term that couples spin and charge degrees
of freedom. The latter is  $\propto a^{2}$. It arises including
spatial derivatives of the fermionic
fields in the representation of (\ref{hamil}) in terms of a low energy field theory.
All of these terms have naive scaling dimension 3 and are usually
neglected. However, one term that bosonize as the second line of (\ref{hboso}) becomes relevant for large enough
$X$ and provides a mechanism for an incommensurate transition, as discussed below.

Explicitly, the effective parameters read $g_{1\perp}=g_{2\perp}=(U-\frac{8X^{2}}{\pi(t-X)})a$
and $g_{\sigma\rho}=\sqrt{2} a^{2} X$.
The forward and umklapp processes are the same as in the Hubbard model,
$g_{3\perp}=g_{4\perp}=Ua$.
The Luttinger liquid parameters ($K_{\rho}$
and $K_{\sigma}$) and the charge and spin wave velocity ($u_{\rho}$
and $u_{\sigma}$) in terms of $g_{i}$ are given by known
expressions \cite{giamarchi}. Neglecting the $g_{\sigma\rho}$ term, the  renormalization-group
(RG) flow diagrams are of KT type. A spin gap
opens when $g_{1\perp}<0$, i.e., when the flow of RG, which takes
place on the separatrix of the KT diagram due to spin SU(2) symmetry, goes
to strong coupling. Therefore, the spin gapped phase appears when $U<U_{s}=\frac{8X^{2}}{\pi(t-X)}$.
As for the behavior of the charge modes, a gap opens when the
$g_{3\perp}$ term becomes relevant. The charge gapped phase takes
place for $U>U_{c}$, with $U_{c}<U_{s}$. The  $g_{\sigma\rho}$ term becomes relevant for $K_{\sigma}<1/2$ ($X>0.6 t$
for $U=0$). In the spin gapped phase the $\cos(\sqrt{8}\phi_{\sigma})$ is
frozen at its mean value. This term could be interpreted as a
chemical potential [$\mu=\frac{2 g_{\sigma\rho}}{(2\pi \alpha)^2} \langle
\cos(\sqrt{8}\phi_{\sigma}) \rangle $] times a charge density operator.
The effects of such a term are known \cite{giamarchi}.
If we start the analysis from a situation where there is also a
charge gap ($\Delta_c$) smaller than the spin one ($\Delta_s$), and
we then increase the value of $X$, the effect of this term is to
close $\Delta_c$, leading to a metallic phase when $\mu > \Delta_c$.
 The effective Fermi level is shifted with respect to the original
one and the system develops incommensurate correlations.
A numerical analysis discussed below shows
that the system has dominant SS correlations.
Thus, this phase can be characterized as incommensurate
singlet superconducting (ICSS).

\begin{figure}
\includegraphics[width=80mm,keepaspectratio,clip]{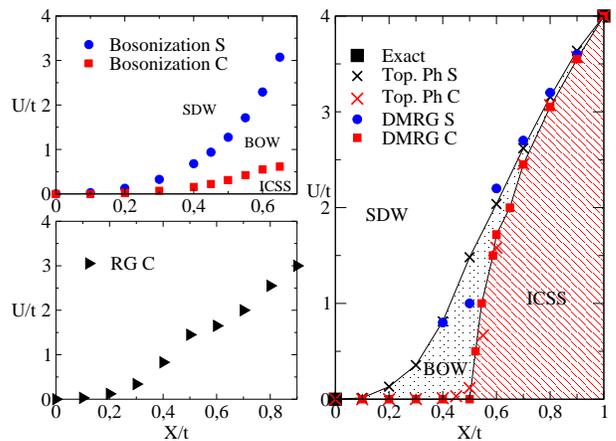}
\caption{(Color online). Phase diagram. Left: Bosonization (top),
and real space renormalization-group (bottom) predictions.
Right: Numerical results as obtained by DMRG (circles-squares) and
topological phase (crosses) methods.}  \label{fig1}
\end{figure}

For a qualitative localization of the boundary transition line
between the insulator and the ICSS phase, we have implemented a
procedure as follows: (i) We start from a parameter regime where the
spin gap is open. (ii) We follow the RG flow up to a length scale
where $|g_{1\perp}|/(\pi U_{s})|\sim1$. (iii) At this point the
$g_{\sigma\rho}$ term is decoupled by a mean field approach
similar to that used by Nersesyan \textit{et al.} to show
incommensurability in the anisotropic zigzag chain
\cite{Nersesyan}. The value of $\langle \cos(\sqrt{8} \phi_{\sigma})
\rangle$ is exactly obtained at the LE point ($K_{\sigma}=1/2$).
(iv) For vanishing $g_{\sigma\rho}$, $\Delta_{c}$ is obtained by rescaling
the problem to the LE point of the charge sector, by using the RG
equations of the sine-Gordon theory. (v) The CIC transition takes
place when $\frac{2 g_{\sigma\rho}}{(2\pi \alpha)^2} \langle
\cos(\sqrt{8}\phi_{\sigma}) \rangle=\Delta_{c}$ \cite{giamarchi}. In  the
top left panel of Fig. \ref{fig1} we show the phase diagram of the model
predicted by this approach. For each value of $X$, there are two
transition points $U_{c}$ and $U_{s}$ corresponding to the charge
and spin transition, respectively. Each phase is characterized by the
gapped modes and the relevant order parameter. For $U>U_{c}$ the
system is an insulator. For $U>U_{s}$, the slowest
decaying correlation functions are the spin-spin ones. The system is in a SDW phase. For
$U_{c}<U<U_{s}$ a fully gapped (spin and charge) phase is developed.
The fields $\phi_{\sigma}$ and $\phi_{\rho}$ are located at the minimum of
the potential, and the translation symmetry is spontaneously broken.
The BOW parameter, defined below, acquires a nonzero value. For
$U<U_{c}$ the charge gap closes and the dominant correlations at
large distances are the SS ones.
While the nature of each phase has been identified, the phase boundaries
predicted by bosonization are not quantitatively valid, particularly for large values of the interactions. In the right panel of Fig. \ref{fig1} we show the phase diagram of the model, as obtained by accurate numerical techniques.
One of them, used to determine the charge transition line,
consists in studying singularities
of single-site entanglement \cite{sup-ins} by means of density-matrix renormalization group (DMRG) \cite{DMRG}. Another method is  based on topological numbers, or jumps of Berry phases \cite{berry}, which was successfully applied to a similar model \cite{japa} (b).
The value of $U_{c}$ ($U_{s}$) is determined in this case by the jump of
the charge (spin) Berry phase. The corresponding values of $U_{c}$
and $U_{s}$ in systems up to $L=14$ sites,
extrapolated to the thermodynamic limit using a parabola
in $1/L^{2}$, are also shown in Fig. \ref{fig1}.

DMRG evaluations of $\Delta_c$ and $\Delta_s$ confirm these predictions. The
charge gap was calculated in \cite{sup-ins} from the definition
$2\Delta_{c}=E_{0}(N+2)+E_{0}(N-2)-2E_{0}(N)$,
$E_{0}(N)$ being the ground-state energy of the chain with $N$ particles.
Similarly, the spin gap is here determined through $\Delta_{s}=E_{0}(S_{z}=1)-E_{0}(S_{z}=0)$,
being $E_{0}(S_{z})$, the ground-state energy of the half-filled
system within the subspace with a given total $S_{z}$. We can see
in Fig. \ref{fig1} that the closing of $\Delta_c$, $\Delta_s$
do not take place simultaneously for small $U$ and $X$. The critical
lines for the closing of both gaps obtained by extrapolations to the
thermodynamic limit are in reasonable quantitative agreement with
the ones determined by the method of the topological phases.

We have verified that the spin transition is of
KT type, calculating the scaling dimensions
of the singlet and triplet operators as described in
\cite{berry}. In order to identify the universality class of the
charge transition, we employed the finite-size
crossing method \cite{cdrs}. The study of the
dependence of $\langle n_{i\uparrow}n_{i\downarrow}\rangle=\partial
e_{L}/\partial U$ on the size $L$ ($e_{L}$ being the ground-state
energy density) provides a location of the critical points in
agreement with the methods discussed above. In addition, the
divergence that develops $\partial e_{L}^{2}/\partial U^{2}$ with
increasing $L$ indicates that the gap exponent $\nu$ remains close
to $1/2$ (the value that can be computed exactly at the point
$X=t$) for $X/t=0.6,\;0.7,\;0.8$ with a possible increase for
$X/t\rightarrow0.5$; below this point, our
numerical analysis suggests that the charge transition becomes of
KT type, with ``$\nu=\infty$''. The estimate $\nu=1/2$ relies
upon the assumption that the dynamic exponent $\zeta$
(through which gap and correlation length  $\xi$ are related,
$\Delta_{c}\propto\xi^{\zeta})$ is still $\zeta=2$, as in
the exactly solvable case $X=t$ \cite{agm}. As already noted in
\cite{sup-ins}, the behavior of $\Delta_{c}\propto L^{-2}$ along the transition line
is consistent with this exponent. We stress that such feature  is in agreement
with the CIC character of the metal-insulator transition \cite{giamarchi}.
Instead, within the metallic phase, the finite-size scaling suggests $\Delta_{c}\propto L^{-1}$, although the data are rather noisy due to incommmensurability.
\begin{figure}
\includegraphics[width=80mm,keepaspectratio,clip]{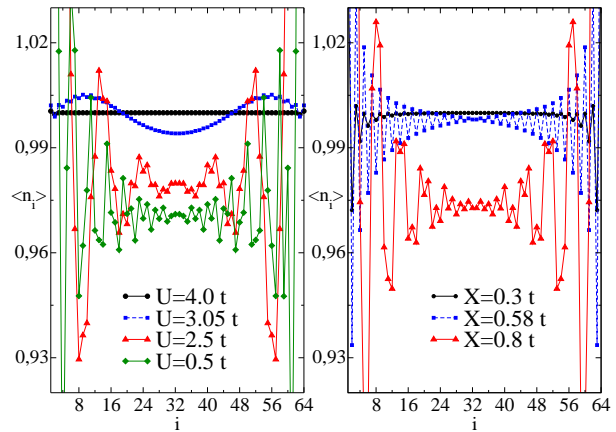}
\caption{(Color online). Charge distribution $\langle n_{i}\rangle$ evaluated by DMRG. Left: $X=0.8 t$. Right: $U=1.5 t$.}
\label{fig2}
\end{figure}

In Fig. \ref{fig2} we show numerical results supporting the
incommensurate character of the metallic phase. We report the
density
distributions in real space for the local charge
density  $n_i=n_{i\uparrow}+n_{i\downarrow}$ in the ground state in an open chain with $L = 64$ sites.
The incommensurate character of the metallic phase manifests itself
also in the behavior of the charge and spin correlation functions,
whose corresponding structure factors show peaks away from the
commensurate reciprocal vector $q=\pi$ (not shown). The left
panel of Fig. \ref{fig2} corresponds to $X= 0.8 t$ as
$U$ is varied. The behavior is similar to the one observed within
the incommensurate phase of the Hubbard model including
next-nearest-neighbor hopping ($t-t'-U$ model) \cite{ttpu}. For
$U>U_c=3.05 t$,  the commensurate charge distribution
characterizing the insulating phase is reached within a few
lattice sites from the edge. The insulator-metal transition
shows up via the appearance of incommensurate modulations
in the charge distribution, whose
wavelength increases within the
metallic phase. The right part of the figure shows the
results obtained by varying $X$ at $U=1.5 t$. Interestingly,
a first modulation appears already for $X_s<X<X_c$ (
$X_s\approx 0.5 t$, and $X_c\approx 0.6 t$). Again, for
$X>X_c$ further incommensurate modulations appear in the LE phase.

\begin{figure}
\includegraphics[width=70mm,keepaspectratio,clip]{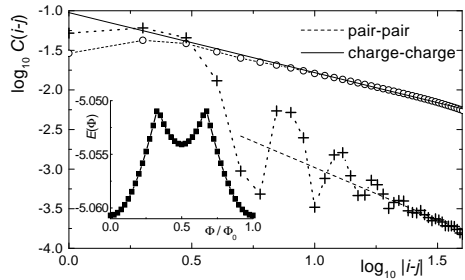}
\caption{Pair-pair and charge-charge correlation functions for $U=t$ and $X=0.8 t$.
Full (dashed) line corresponds to a power law with exponent $1/K_{\rho}$ ($K_{\rho}$).
The inset shows the ground state energy as a function of an applied magnetic flux.}
\label{fig3}
\end{figure}

Within the charge sector $U<U_{c}$, the dominating correlations at large distance
are superconducting pair-pair ones if the correlation exponent $K_{\rho}>1$ or
charge-charge ones otherwise.
We calculated $K_{\rho}$ employing
the methodology described in \cite{tll}. This study casts extrapolated
values $K_{\rho} \sim 1.3$ for $U=0$ and $X=0.8 t$. To provide stronger
evidence for the SS character of the incommensurate phase, we have
calculated on-site pairing correlations $\langle P_{i}^{\dagger }P_{j}\rangle $ with
$P_{i}=c_{i\uparrow }^{\dagger }c_{i\downarrow }^{\dagger }$ and
charge-charge correlations $|\langle n_{i}n_{j}\rangle -\langle n_{i}\rangle \langle
n_{j}\rangle |$  in an open chain with 100 sites and using the sites 30 to 70
to avoid boundary effects.
The results are displayed in Fig. \ref{fig3}.
A fitting of the pairing correlations at distances between 8 and 40 sites
gives $K_{\rho}=1.32 \pm 0.01$.
This value is also consistent with the long distance behavior of the
charge-charge correlations.
The inset also shows the tendency of the system to show the anomalous flux
quantization characteristic of superconductivity \cite{flux}, which is
more pronounced as the size of the system increases.

An additional argument suggesting superconducting correlations within
this phase is provided by the real space renormalization-group method,
used before for the standard Hubbard model \cite{HIR}.
Different from that case, the recursive equations
for the renormalized parameters in the positive $U$ regime, depending on $X$ and $U$,
exhibit three different fixed points for the $n$th step renormalized Coulomb
interaction $U^{(n)}$ in the large $n$ limit: $U^{(n)}>0$ for $U>U_{rc}$,
$U^{(n)}=0$ for $U=U_{rc}$, and $U^{(n)}<0$ for $U<U_{rc}$. In the latter case,
the effective Coulomb interaction becomes attractive. In the bottom left insert
of Fig. \ref{fig1} $U_{rc}$ obtained in this way is reported.

To support the bosonization predictions, which characterize the intermediate phase as a BOW,
we have evaluated with DMRG the BOW order parameter $O_{BOW}=[\sum_{i,\sigma}(-1)^{i}\langle c_{i+1\sigma}^{\dagger}c_{i\sigma}+H.c.\rangle]/(L-1)$ in chains with open boundary conditions,
following the same procedure as Manmana \textit{et al.} for the ionic Hubbard model \cite{manma}
in chains up to 400 sites.  In spite of the large systems used, finite-size
effects are still important and do not allow an accurate extrapolation. In any
case, the qualitative behavior of our results (not shown) is similar to that found by Manmana
\textit{et al.} showing a clear maximum inside the BOW phase, an abrupt fall for $U \sim U_{c}$ as
the system enters the SS phase and a slower decay for larger $U\sim U_{s}$, which for finite
systems extends inside the SDW phase.

To conclude, we have presented compelling evidence, based on
bosonization as well as on other analytical and numerical
techniques, of the existence of a narrow bond-ordered wave phase
and a transition to an unconventional incommensurate metallic one with
dominant singlet superconducting correlations in the
phase diagram of the $U-X$ model. The appearance of superconductivity in a model with repulsive on-site interactions at half filling, and of incommensurate correlations induced by interaction are both unusual features. Their emergence can be understood from the structure of the exactly
solvable case $X=t$. There the number $N_d$ of doubly occupied sites
(doublons) becomes a conserved quantity; holes and doublons play an
identical role regarding the kinetic energy $\epsilon(k_F)$, which can be mapped
into that of a spinless fermion system, with Fermi momentum $k_F$.
The competition of $\epsilon(k_F)$ and $U N_d$ fixes the Fermi level
of the resulting effective model.
The presence of doublons in the ground state  ($U< 4 t$)
simultaneously drives the spinless fermions away from half-filling
($k_F\neq \pi$), and switches on the doublons role in the kinetic
energy. The latter ceases to be identical to that of holes as soon as $X\neq t$, generating incommensurability within the system. Moreover superconducting correlations can dominate away from half-filling \cite{japa}.  Thus, a nonvanishing number of doublons provides the scenario for both incommensurability and superconductivity for $X\lesssim t$.

We thank D. Cabra for useful discussions. We acknowledge support from
PICT's No. 03-11609, No. 03-12742, and No. 05-33775 of ANPCyT and PIP's No. 5254 and No. 5306 of CONICET,
Argentina, No. FIS2006-08533-C03-02, and the ``Ramon y Cajal'' program from MCEyC of Spain,
Angelo Della Riccia Foundation, and PRIN 2005021773 Italy.

\end{document}